\documentclass[journal=jacsat,manuscript=article]{achemso}
\setkeys{acs}{articletitle = true}
\usepackage{expl3}
\usepackage{amsfonts}
\usepackage{graphicx, type1cm, lettrine}

\usepackage[version=3]{mhchem} 
\usepackage{color}
\usepackage[colorlinks=true, letterpaper=true, pdfstartview=FitV, linkcolor=blue, citecolor=blue, urlcolor=blue]{hyperref}

\usepackage{mathrsfs}
\usepackage{amsmath}
\usepackage{bm}
\usepackage{amssymb}
\usepackage{xspace}
\usepackage{epstopdf}
\usepackage{dcolumn}
\usepackage{longtable}
\usepackage{multirow}

\author{Cong Chen}
\affiliation{Department of Physics, Key Laboratory of Micro-nano Measurement-Manipulation and Physics (Ministry of Education), Beihang University, Beijing 100191, China}

\author{Zefeng Su}
\affiliation{Department of Physics, Key Laboratory of Micro-nano Measurement-Manipulation and Physics (Ministry of Education), Beihang University, Beijing 100191, China}

\author{Xiaoming Zhang}
\affiliation{School of Materials Science and Engineering, Hebei University of Technology, Tianjin 300130, China}

\author{Ziyu Chen}
\email{chenzy@buaa.edu.cn}
\affiliation{Department of Physics, Key Laboratory of Micro-nano Measurement-Manipulation and Physics (Ministry of Education), Beihang University, Beijing 100191, China}

\author{Xian-Lei Sheng}
\email{xlsheng@buaa.edu.cn}
\affiliation{Department of Physics, Key Laboratory of Micro-nano Measurement-Manipulation and Physics (Ministry of Education), Beihang University, Beijing 100191, China}
\alsoaffiliation{Research Laboratory for Quantum Materials, Singapore University of Technology and Design, Singapore 487372, Singapore}

\title{From Multiple Nodal Chain to Dirac/Weyl Semimetal and Topological Insulator in Ternary Hexagonal Materials}
\abbreviations{}

\begin{document}
\begin{abstract}
Dirac semimetal (DSM) hosts four-fold degenerate isolated band-crossing points with linear dispersion, around which the quasiparticles resemble the relativistic Dirac Fermions. 
It can be described by a $4\times 4$ massless Dirac Hamiltonian which can be decomposed into a pair of Weyl points or gaped into an insulator. Thus, crystal symmetry is critical to guarantee the stable existence. On the contrary, by breaking crystal symmetry, a DSM may transform into a Weyl semimetal (WSM) or a topological insulator (TI). Here, by taking hexagonal LiAuSe as an example, we find that it is a starfruit shaped multiple nodal chain semimetal in the absence of spin-orbit coupling (SOC). In the presence of SOC, it is an \emph{ideal} DSM naturally with the Dirac points locating at Fermi level exactly, and it would transform into WSM phase by introducing external Zeeman field or by magnetic doping with rare-earth atom Sm. It could also tranform into TI state by breaking rotational symmetry. Our studies show that DSM is a cirtical point for topological phase transition, and the conclusion can apply to most of the DSM materials, not limited to the hexagonal material LiAuSe.
\end{abstract}

\newpage

\section{Introduction}
\lettrine[lines=2, findent=3pt, nindent=0pt]{T}{}opological semimetals have been attracting wide-range of research interest in condensed matter physics~\cite{RevModPhys.88.035005,Dai:2016bu,Zhao2013c}, partly because they give access to new quantum phenomena and offer a platform to investigate the intriguing properties of high-energy particles~\cite{Volovik2003}.  Weyl semimetal (WSM) has linear despersion around  two-fold degenerate band-crossing points~\cite{WanXG_Weyl,1367-2630-9-9-356,Balents_2011Weyl,WengHM_2015TaAs,Huang:2015ic,Haijun2014PRL,Lv2015,Xu613,ShengQAHE,YuZM2016}, and the Weyl point possesses a definite chirality of $\pm 1$, around which the quasiparticle excitation is anolog of Weyl fermions~\cite{Volovik2003,NIELSEN1983389}.  Furthermore, it could be viewed as the monopole of Berry flux in momentum space, and the two Weyl points with opposite chiralities correspond to the source and drain respectively~\cite{RevModPhys.82.1959}. The surface states of a WSM are Fermi arcs, which are open segments of Fermi surface connecting the projections of  Weyl points of opposite chiralities on the two-dimentional (2D) surfaces~\cite{WanXG_Weyl}. An isolated Weyl point is a separate band-crossing node of double degeneracy, which can be described by the Hamiltonian $H=v_F\bm{\sigma} \cdot \bm{k}$ in low-energy physics, where $v_F$ is the Fermi velocity, $\bm{\sigma}$ is the vector of Pauli matrices, and $\bm{k}$ is the momentum measured from the band-crossing point. In materials with both time reversal symmetry $\mathcal{T}$ and inversion symmetry $\mathcal{P}$, all the bands are double degenerate and there should exist band-crossing points of four-fold degeneracy with linear dispersion around the nodal points. Such a material is known as 3D  Dirac semimetal (DSM)~\cite{PhysRevLett.108.140405,PhysRevB.85.195320,Liu864,PhysRevB.88.125427,Liu2014Exp_Cd3As2,Neupane:2014kc,PhysRevLett.113.027603,ChenC2017,Guan2017PRM,ZhangXM2017}. Around these points, the low-energy physics is described by a $4\times 4$ massless Dirac Hamiltonian. In Weyl representation, a fourfold Dirac matrix can be decomposed into a block diagonalised matrix consisting of two copies of Weyl nodes with opposite chiralities. Hence, there are two kinds of Fermi arcs: (i)  open segments connecting two different Dirac points; (ii) closed cicles beginning and ending at the same Dirac point. The former are similar to the case in WSM but occur in pairs. The later may shink into a point for each Fermi arc.  

The Weyl Hamiltonian is quite robust against any perturbation~\cite{Yang2016}. A small perturbation expanded around the Weyl point takes the general form of $\delta_0I_{2\times2}+\bm{\delta}\cdot\bm{\sigma}$, where $\delta_0$ and $\bm{\delta}$ are Taylor expansions in wave vector $\bm{k}$, and $I_{2\times2}$ is a $2\times2$ identity matrix. The first term only shifts the Weyl point in energy, while the second term only shifts its location in momentum space. Thus, the Weyl points cannot be removed by any small perturbations.
However, the DSM phase is not stable against perturbations because additional symmetry is necessary~\cite{PhysRevB.85.195320,PhysRevB.88.125427,ChenC2017}. If $\mathcal{T}$ or $\mathcal{P}$ symmetry was broken, DSM may be transformed into WSM~\cite{DuYP2015SR,Sheng2017JPCL}. If external strain was introduced, the Dirac point could be gaped and the DSM would be tranformed into a topological insulator~\cite{ZhangHJ2011,PhysRevB.96.075112,Guan2017}. The above analysis is well known for decades, but the transformation process is rarely observed in real materials.

In this paper, we will investigate the topological phase transition  by revealing a new class of DSM  in ternary compound LiAuSe family, which has been predicted as a TI under rotational symmetry breaking~\cite{ZhangHJ2011}. LiAuSe has a layered honeycomb lattice of AuSe and sandwiched by Li atoms with triangular lattice, possessing $P6_3/mmc$ (or $D_{6h}^{4}$) space group.  Band inversion exists in this material, and it is a starfruit shaped multiple nodal chain semimetal in the absence of spin-orbit coupling (SOC), while in the presence of SOC, the nodal chain breaks and the system becomes a Dirac semimetal with a pair of Dirac nodes along $k_z$ line protected by $C_6$ rotational symmetry. Please note that it is the first time here to report starfruit shaped multiple nodal chain metal phase and the phase transition to Dirac semimetal, which is different from other nodal line or nodal chain semimetal~\cite{WengHM2015carbon,Chen2015,YuR2015PRL,ZhaoJZ2016,XuQN2017,YuR2017,WangSS2017,ZhangXM2017jpcl,ZhangTT2017,LiS2017,DuYP2017}
By introducing external magnetic field or magnetic atoms doping, it transforms into WSM phase. By breaking rotational symmetry, it transforms into TI phase. Furthermore, we constructe a low-energy $k\cdot p$ model to describe the topological phase transitions.

\begin{figure}[b!]
\centerline{\includegraphics[clip,width=15cm]{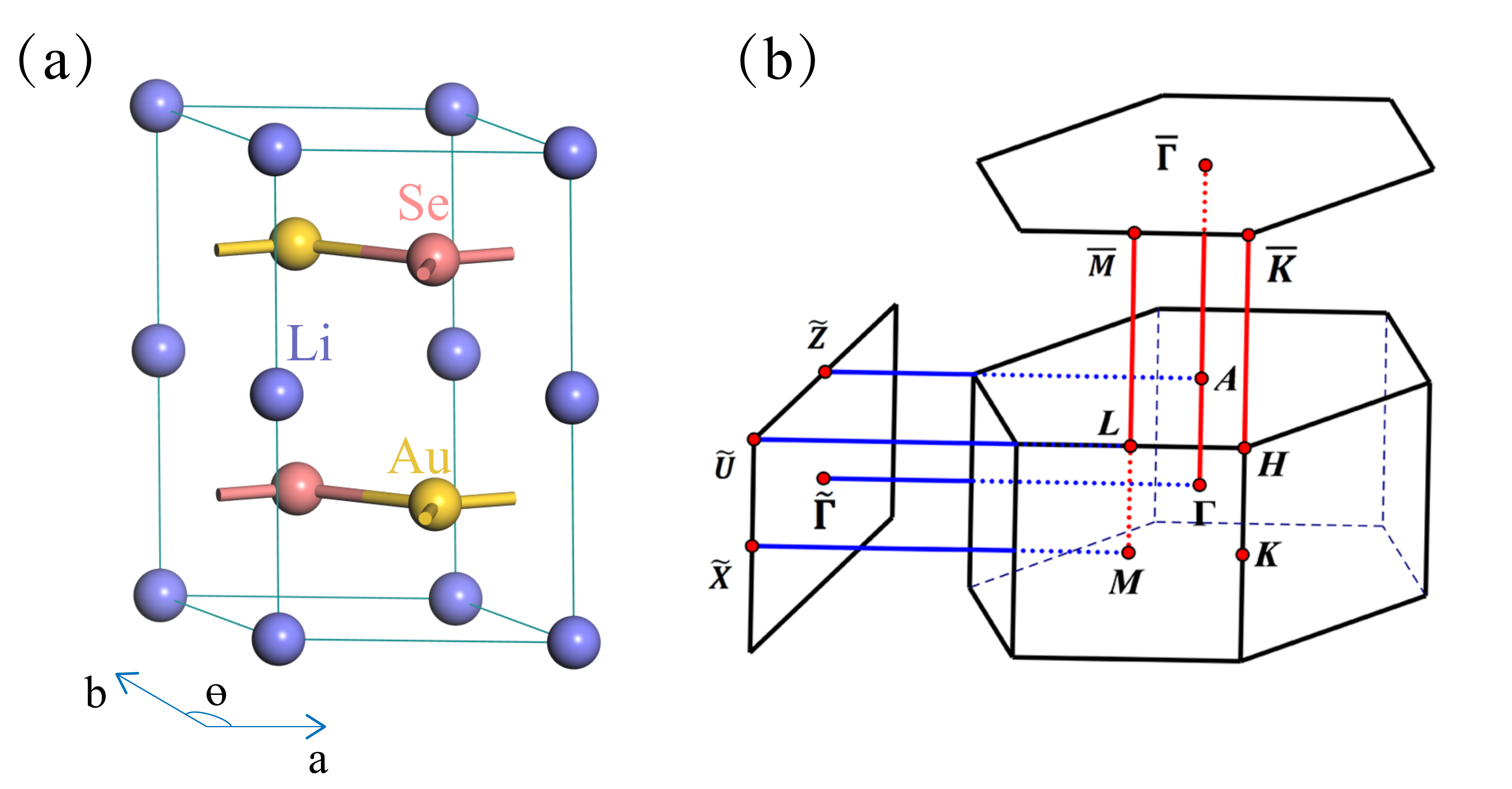}}
\caption{(a) Crystal structure of LiAuSe. (b) Corresponding Brillouin zone (BZ), as well as the projected surface BZ of (0001) and (0100) surfaces.}
\label{fig1}
\end{figure}

\begin{figure}[b!]
\centerline{\includegraphics[clip,width=15cm]{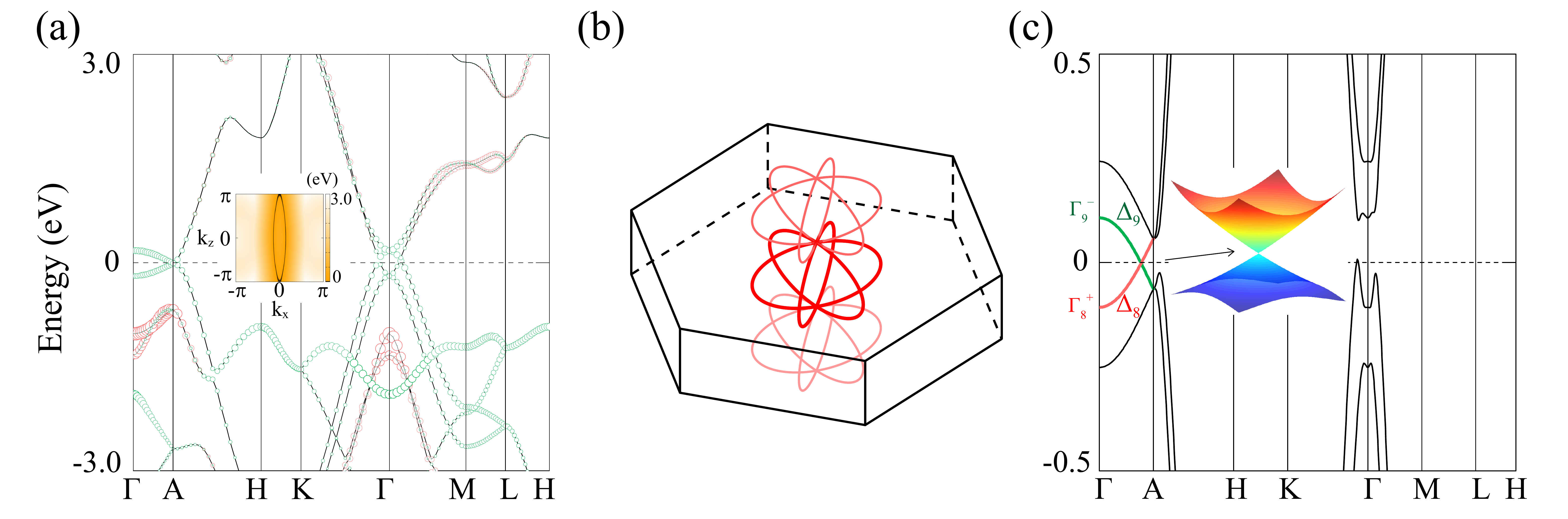}}
\caption{(a) Band structure of LiAuSe without SOC. The colored circles indicate the weight of orbital projection onto Au-$s$ (red) and Se-$p$ (green) orbitals. The inserted figure is the nodal loop on $\Gamma$-M-A plane in BZ. (b) Further considering $C_6$ rotational symmetry and translation symmetry along $k_z$ direction, these nodal loops form a starfruit-like nodal line in the first BZ and a multiple nodal chain in momentum space. (c) The band structure with SOC and the irreducible representation of the two crossing bands along $\Gamma$-A around Fermi level. The inserted 3D band structure is the corresponding Dirac cone.}
\label{fig2}
\end{figure}

\section{Computational Methods and Materials}

All the first-principles calculations  are based on density-functional theory (DFT) as implemented in Vienna \textit{ab initio} simulation package (VASP) \cite{VASP1,VASP2} with the projector augmented wave (PAW) method \cite{paw}. The generalized gradient approximation (GGA) with Perdew-Burke-Ernzerhof (PBE) \cite{GGAPBE} realization were adopted for the exchange-correlation potential. The plane-wave cutoff energy was taken as 500 eV. The Monkhorst-Pack $k$-point mesh~\cite{PhysRevB.13.5188} of size $10\times10\times 10$ was used for Brillouin zone sampling.  The crystal structures were optimized until the ramanent forces on the ions were less than 0.01 eV/\AA. From the DFT results, we constructed the maximally localized Wannier functions (MLWF)~\cite{Wannier90T,Wu2017} for Au-$s$ and Se-$p$ orbitals, and effective model Hamiltonian for bulk and semi-infinite layer were built to investigate the topological invariants and the surface states~\cite{Sheng_TlN,ShengQAHE,Sheng2017JPCL,CHEN20173337}.

We take the ternary compound LiAuSe as an example, which can be viewed as a stuffed honeycomb structure, where Au and Se atoms form a honeycomb lattice, stacking layer by layer along the z dimension, and Li atoms are inserted between these layers. Such a structure possess $P6_3/mmc$ (or $D_{6h}^{4}$) space group symmetry (No. 194). The crystal structure and corresponding Brillouin zone (BZ), as well as the projected surface BZ of (0001) and (0100) surfaces are shown in Figs.~\ref{fig1} (a) and ~\ref{fig1} (b).  We take the relaxed lattice parameter as $a=b=4.42$~\AA\ and $c=7.38$~\AA, in good agreement with the previous result~\cite{ZhangHJ2011}.

\section{Results and Discussions}

\subsection{In the absence of SOC: nodal chain semimetal}
We first consider the band structure of LiAuSe in the absence of spin-orbit coupling (SOC), as shown in Fig.~\ref{fig2} (a). From the orbital projected band structure, one can observe that LiAuSe is a semimetal and the low-energy states near the Fermi level are mainly from the Se-$4p$ orbitals and Au-$6s$ orbitals. Furthermore,  around $\Gamma$ and $A$ points, Au-$6s$ states are lower than Se-$4p$ in energy by about 1.0 eV, while around $M$ and $L$ points, Au-$6s$ states are energetically higher than Se-$4p$ states, showing  a band inversion character~\cite{Sheng_TlN,Sheng2017JPCL,PhysRevLett.112.21683}. The band inversion character is similar to HgTe, where Hg-$6s$ states are energetically lower than Te-$5p$ states around $\Gamma$ point, while for trivial insulator GaAs, the cation-$s$ states are energetically higher than anion-$p$ states in the whole BZ. 

There are several nontrivial band features near Fermi energy. Fig.~\ref{fig2}(a) shows two linear band-crossing points at $A$ and along the $\Gamma$-$M$ path. Interestingly, these two points are not isolated. After carefully scanning, we find that they are located on a nodal loop in the $\Gamma$-$M$-$A$ plane. Considering the $C_6$ rotational symmetry, there are three such nodal loops in the first BZ, forming starfruit-like nodal loops. Furthermore considering translational symmetry along $k_z$ axis, these nodal loops form a multiple nodal chain, as shown in Fig.~\ref{fig2} (b).

\subsection{In the presence of SOC: intrinsic Dirac semimetal}

\begin{figure}[b!]
\centerline{\includegraphics[clip,width=15cm]{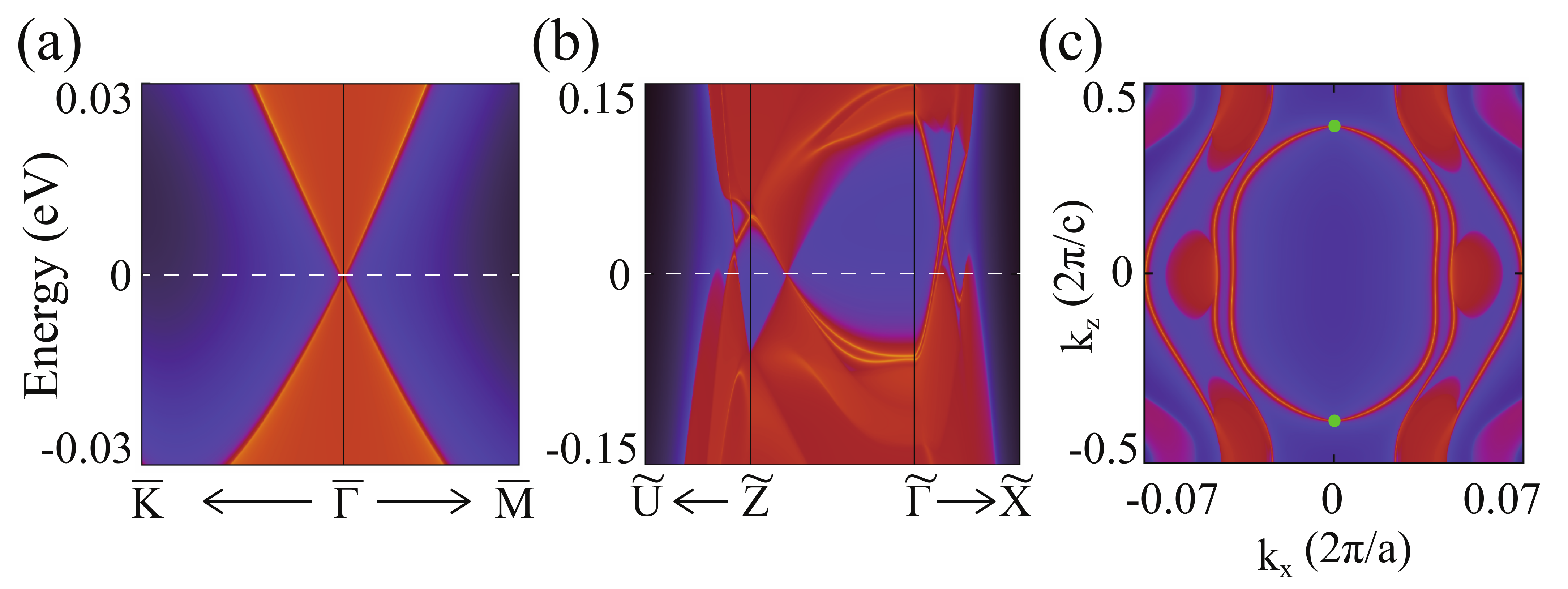}}
\caption{(a) Projected spectrum on (0001) surface for LiAuSe, showing solid Dirac-cone surface states owing to the existence of Dirac points in the bulk band structure. (b) Projected spectrum on (0100) surface for LiAuSe, showing (c) surface Fermi arcs connecting a pair of projected Dirac points (green dots).}
\label{fig3}
\end{figure}

At $\Gamma$ point, the electronic states around Fermi level is mostly from Se-$4p_{x,y}$ states.   After considering SOC, there is a band gap (about 0.2 eV) openned at $\Gamma$ point, indicating the orbital splitting of Se-$4p$ under SOC. However, there is still a band crossing along $\Gamma$-$A$ line [Fig.~\ref{fig2} (c)]. Since both time-reversal and inversion symmetries are present, it is of fourfold degeneracy at each crossing point, around which the band dispersions are linearized. Therefore, LiAuSe is an \emph{ideal} 3D Dirac semimetal with a pair of Dirac points at $\Gamma$-$A$ line. For such a band structure, the Fermi level is only a pair of two points at $k_z=\pm 0.388~2\pi/c$, coined as "Fermi points". 

To confirm the existence of the Dirac nodes in LiAuSe, we calculated the irreducible representations for each band. The two crossing bands belong to different irreducible representations, $\Gamma_8$ ($E_{5/2}$) and $\Gamma_9$ ($E_{3/2}$), and with different parities, indicating that the Dirac node is unavoidable. It is protected by the three-fold rotational symmetry. Breaking this symmetry would introduce interaction between the two bands and open a gap inbetween, making the system into a topological insulator with $Z_2=1$ due to the band inversion, which has been confirmed by our first-principles calculations. For the $P6_3/mmc$ structure LiAuSe, inversion symmetry is preserved. Thus, we can calculate the parity product at the eight  time reversal invariant momentum (TRIM) points to confirm the nontrivial band topology~\cite{Fu2007}. We find that this product is positive at $\Gamma$ point and negative at other TRIMs, in accordance with our analysis of the band inversion, which leads to a Dirac semimetal phase for the initial structure, and a strong TI with $Z_2$ indices (1;000) after breaking rotational symmetry.

It has been shown that in Na$_3$Bi and Cd$_3$As$_2$ Dirac points in the bulk may lead to Fermi arc surface states at the sample surface. In Fig.~\ref{fig3} (a) we plot the surface spectrum of the (0001) surface. Since the two Dirac points locate at $\Gamma$-$A$ line, one observes that there appears a solid Dirac cone, which is the projection of the two Dirac points in the bulk. In Figs.~\ref{fig3} (b) and ~\ref{fig3} (c) we plot the surface spectrum on the (0100) surface. Since this plane is parallel to the $\Gamma$-$A$ line, it is clear to see the Dirac cone from each Dirac point in the bulk and two Fermi arcs connecting the two Dirac points. Since each Dirac point consists of a pair of Weyl points, there are two Fermi arcs. To see each Fermi arc separately, it is necessary to break band inversion symmetry as discussed in the following subsection.

\subsection{Dirac semimetal to Weyl semimetal phase transition by Zeeman field}

\begin{figure}[b!]
\centerline{\includegraphics[clip,width=15cm]{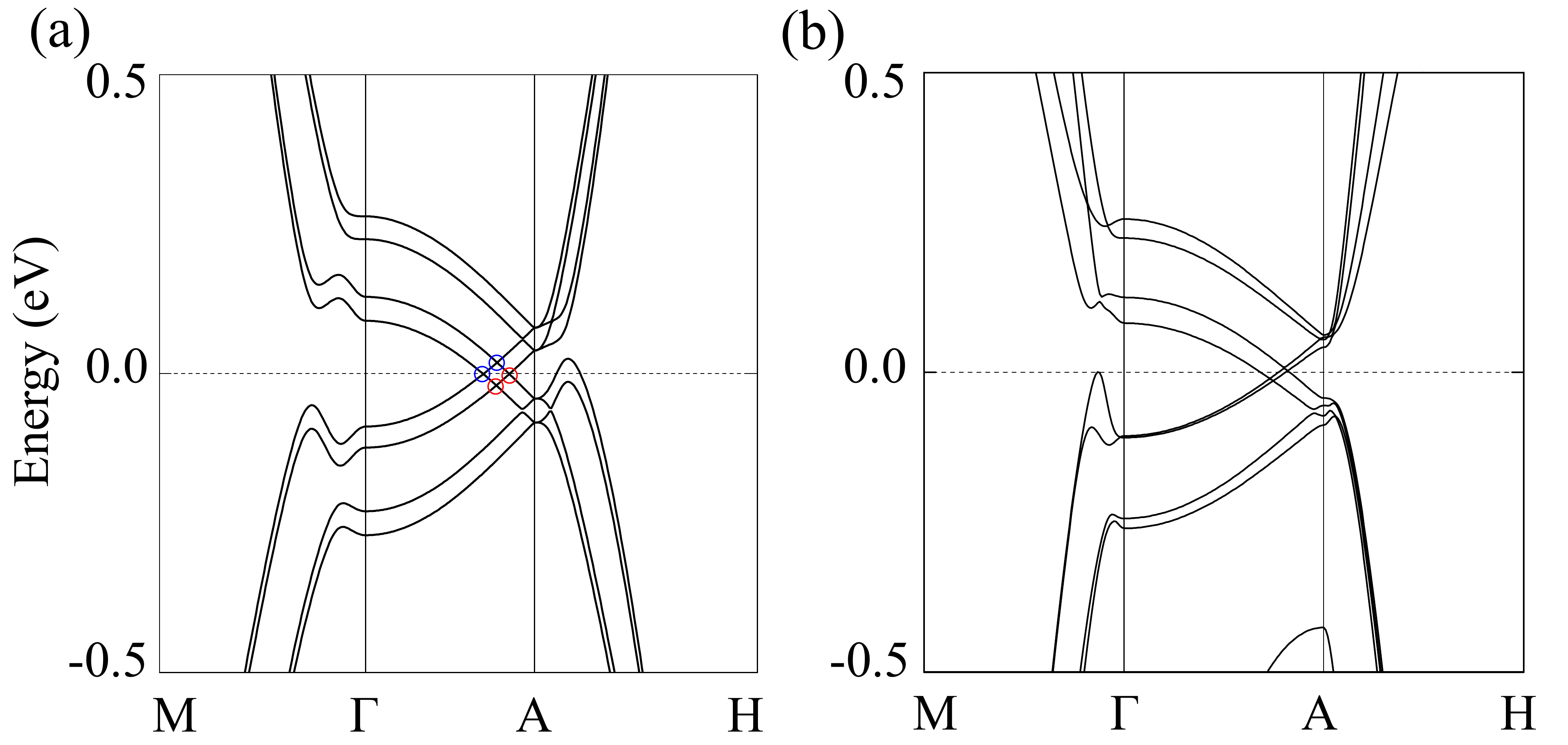}}
\caption{(a) Band structure of LiAuSe with external field $h=0.02$ eV. The double degeneracy was split for the breaking of time-reversal symmetry. (b) Band structure of Li$_{0.5}$Sm$_{0.5}$AuSe. In this case, half Li atoms in LiAuSe were replaced by magnetic Sm atoms. Thus, double degeneracy was also split due to magnetic doping.}
\label{fig4}
\end{figure}

\begin{figure}[b!]
\centerline{\includegraphics[clip,width=15cm]{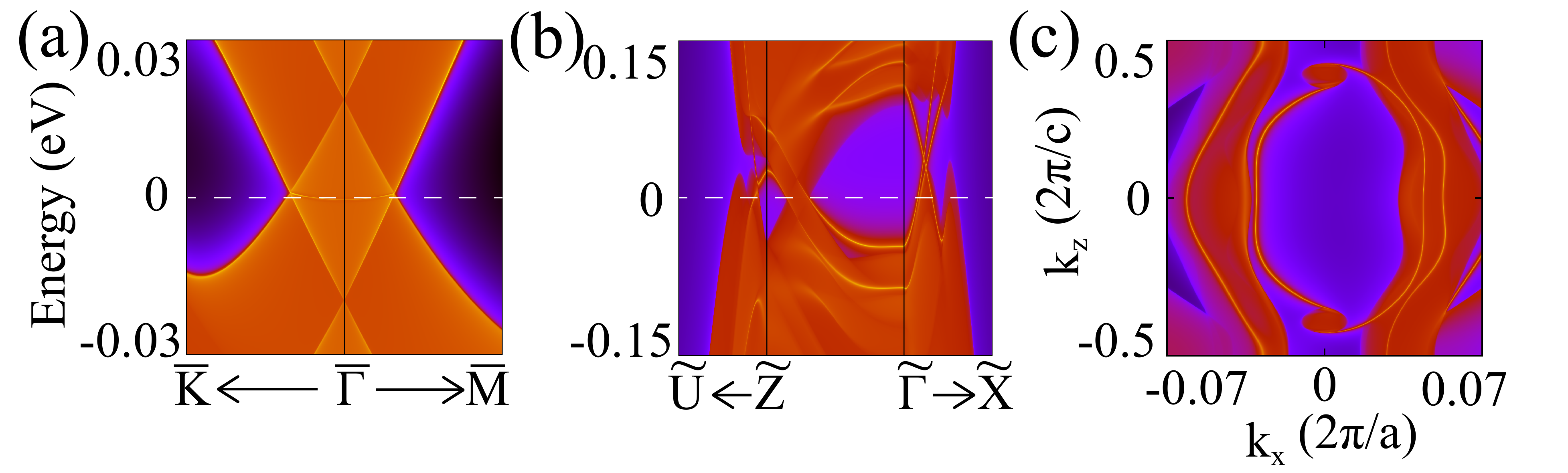}}
\caption{Projected spectrum of LiAuSe (a) on (0001) surface and (b) on (0100) surface. (c) Surface Fermi arcs connecting a pair of projected Weyl points on (0001) surface.}
\label{fig5}
\end{figure}

The Dirac nodes in LiAuSe are of fourfold degeneracy, and the disperssions along any direction are doubly degenetated due to the coexistence of inversion and time-reversal symmetries. Upon breaking the time reversal or inversion symmetry, a Dirac point would split into a pair of separated Weyl nodes. Since there is a three-fold rotational axis, the Dirac points cannot be split by breaking inversion symmetry, such as CaAgBi with wurtzite structure~\cite{ChenC2017}. Thus, we only consider breaking time reversal symmetry in the system. In general, such a system can be induced by magnetic doping as in diluted magnetic semiconductors, or by an external magnetic field. The rare earth element Eu doped compounds of hexagonal structure, e.g. EuAgBi and EuAuSb,  indeed exist in nature, and the long-range magnetic configuration in some of them has been confirmed in experiments. Therefore, a certain degree doping of Eu in the hexagonal compound could be expected as a WSM, which has been confirmed by first-principles calculations in Eu$_{0.5}$Ba$_{0.5}$AgBi~\cite{DuYP2015SR}.

Here, we firstly adopt another way of realizing WSM in LiAuSe system, by applying external field in the model Hamiltonian. Once external magnetic field applied or FM order is achieved in the system, the simplest low-energy effective Hamiltonian is described as
\begin{equation}
 \begin{split}
 \label{WSM}
 \mathcal{H}=\sum_{ij\alpha\beta}t_{ij}^{\alpha\beta}C_{i\alpha}^{\dagger}C_{j\beta} +\sum_{i}h\sigma_{\alpha\beta}^z C_{i\alpha}^{\dagger}C_{i\beta}
 \end{split}
 \end{equation} 
where $C_{i\alpha}^{\dagger}$ creates an electron with spin $\alpha$ on the site (and orbital included) $i$ of the 3D hexagonal lattice, $t_{ij}^{\alpha\beta}$ is the hopping parameter, h is the field strength, and $\sigma_{\alpha\beta}^z$ is the element of Pauli matrix $\sigma^z$. The first term is the tight-binding Hamiltonian with SOC included which can be fitted by first-principles calculations, and the second term is the Zeeman term with the external field along $z$ direction and the strength $h=20$ meV. After applying this term, the time-reversal symmetry is broken, and the doubly degenerated bands are split into single ones [see Fig.~\ref{fig4}(a)]. Consequently, the Dirac point on $\Gamma$-$A$ was split into a pair of Weyl points with opposite chiralities. Since the external field is along $z$-axis, the rotation symmetry is still preserved. Therefore, the Weyl points are still on the $k_z$-axis as well as Dirac points. In addition, another two Weyl points emerge above or below the pair of expected Weyl nodes by 20 meV.  The Zeeman field strength could also take other value, such as 40 meV, it is nothing more than change of the band splitting range. Furthermore, we find that band degeneracy could also be removed in magnetic atom doped LiAuSe system. In order to achieve Weyl semimetal phase in LiAuSe by magnetic atoms doping, there are two key points: (i) the doping atoms could provide an effective magnetic field; (ii) the band coupling between the magnetic atoms and parent material should week enough to maintain the band crossings and keep them around Fermi level. Then, we have calculated many structures by doping different magnetic atoms, and find that Sm (or Ir, Os) dopped LiAuSe system satisfy the expectation [See Fig.~\ref{fig4} (b)].  In both cases, the Dirac point split into Weyl points, and all the Weyl points are very close to the Fermi level, indicating that they could be observed by experiments. 

Figure~\ref{fig5} (a) shows the projected surface band structure on (0001) plane, on which the pair of Weyl nodes are projected at $\Gamma$ point, so that no Fermi arc can be observed. On surface (0100), which is parallel to $z$-axis, the Weyl nodes are separated on $k_z$-axis. Figure~\ref{fig5} (b) shows the projected band structure on this surface, where the topological nontrivial surface states can be observed clearly. The Fermi arcs are separated, and they are connecting the Weyl points in pair [see Fig.~\ref{fig5} (c)],  corresponding to the two arcs in Fig~\ref{fig3}(c) connecting the two Dirac points. 

\begin{figure}[b!]
\centerline{\includegraphics[clip,width=15cm]{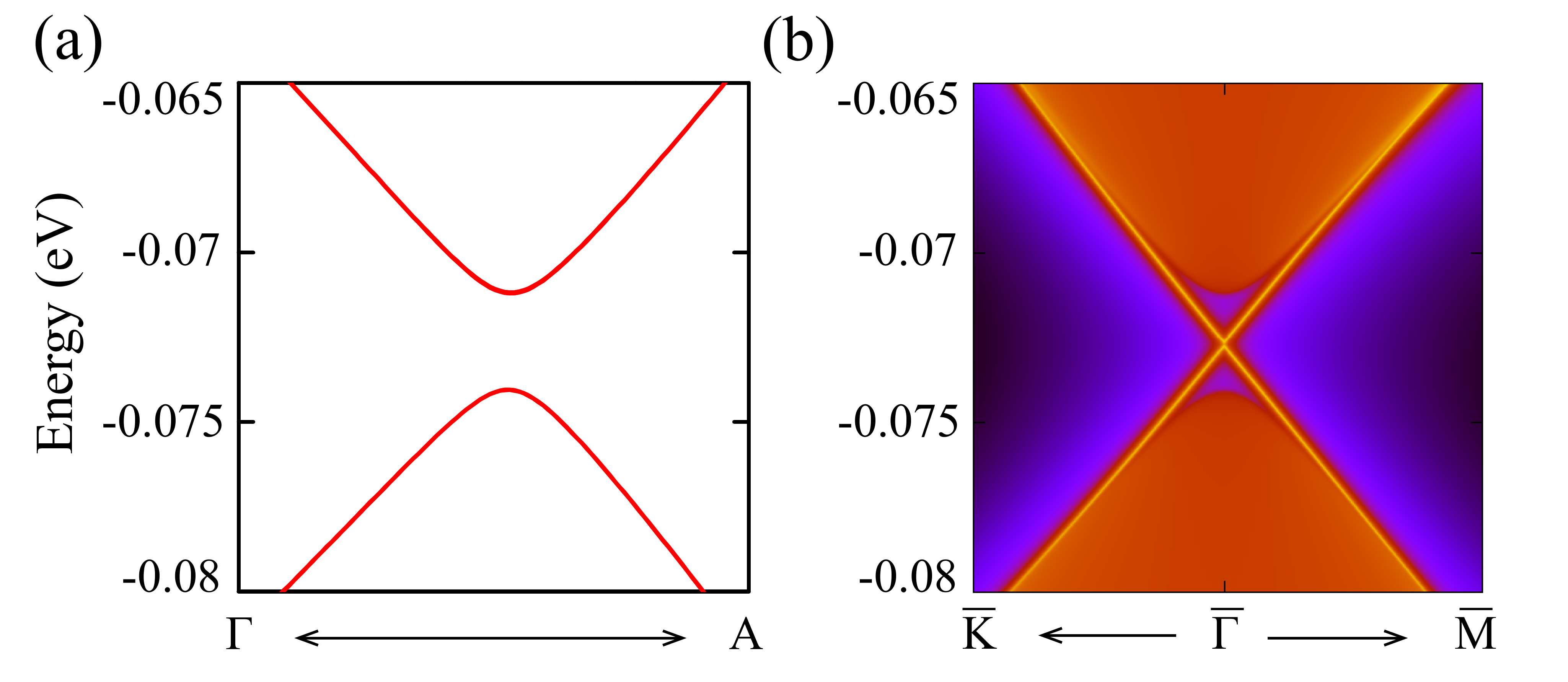}}
\caption{(a) Band structures of LiAuSe with SOC after breaking $C_6$ symmetry by changing the angle of vector lattice $\bm{a}$ and $\bm{b}$. (b) The corresponding projected spectrum on (001) surface. }
\label{fig6}
\end{figure}

\subsection{Dirac semimetal to topological Insulator phase transition by breaking rotational symmetry}


As mentioned above, the accidental Dirac points are caused by band inversion and protected by rotational symmetry. Therefore, the Dirac points can be removed by breaking the rotational symmetry. In the case of LiAuSe, we break this symmetry by applying an in-plane compression, which decreases the angle $\theta$ between $a$ and $b$ axis as shown in Fig.~\ref{fig1}(a). As a result,  the system becomes a true insulator. A moderate compression (${114}^{\circ} \leq \theta \leq {120}^{\circ}$) could open a gap in the bulk and transform  the system into TI phase. Since band inversion is still kept by checking the orbital projected band structure, it is a strong TI with topological invariant (1;000), which is confirmed by  the Wilson loop method calculations and the parity products of occupied states at TRIM points. Fig.~\ref{fig6} (a) shows the band structure with SOC included when $\theta$=${114}^{\circ}$. It is clear that the Dirac type band crossing is removed and  a band gap of about 3 meV openned. For a strong TI, topological nontrivial surface states should  exist. The projected spectrum on (0001) surface  with a single Dirac point at $\Gamma$ in the energy gap is shown in Fig.~\ref{fig6} (b).

\subsection{Band structures of other materials in the family}

\begin{figure}[b!]
\centerline{\includegraphics[clip,width=15cm]{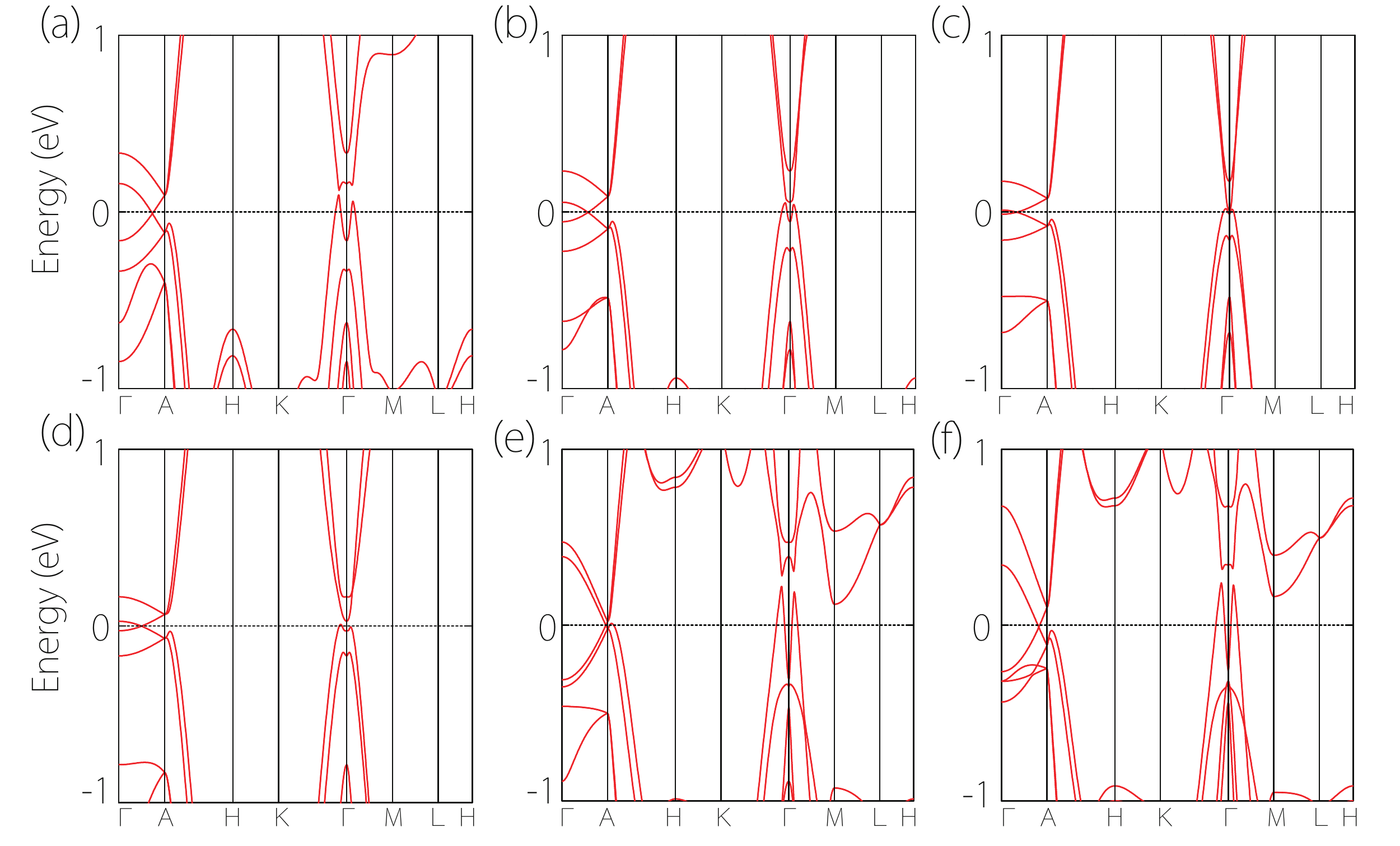}}
\caption{Band structure of some other materials in LiAuSe family with SOC: (a) LiAuTe, (b) NaAuTe, (c) KAuTe, (d) NaAuSe, (e) BaAuAs, and (f) BaAuSb}
\label{fig7}
\end{figure}

Figure~\ref{fig7} shows the band structure of some other materials of LiAuSe family with SOC: LiAuTe, NaAuTe, KAuTe, NaAuSe, BaAuAs, and BaAuSb. These materials are of the same family and with the same symmetry, showing similar characters in their electronic structures. Thus, we state that similar topological phase transition could be expected for other Dirac materials.

\subsection{Effective $k\cdot p$ model}

To further understand the band structure of the Dirac nodes in LiAuSe, a low-energy effective Hamiltonian is derived. The first-principles calculations show that the irreducible representations of two crossing bands at $\Gamma$ are $\Gamma_8^{+}$ and $\Gamma_9^{-}$ (see Fig.~\ref{fig2}). The symbol $+$ and $-$ indicate the parity of the states. From the orbital projected band structure, we know that $\Gamma_8^{+}$ and $\Gamma_9^{-}$ are mainly from Au-$s$ and Se-$p$ states. By including SOC, spin and orbital angular momentum are coupled. Furthermore, under $D_{6h}$ symmetry, the band inversion and the Dirac points can be described using  the four states $| S_{1/2,\pm1/2}\rangle$ and $P_{3/2, \pm3/2}$. Therefore, an effective  $k\cdot p$ Hamiltonian using these four basis functions (basis order: $|i \uparrow \rangle$, $| (x+iy)\uparrow \rangle$, $|i\downarrow \rangle$,  $| (x-iy)\downarrow \rangle$) can be constructed under the constraint of  $D_{6h}$ symmetry, which can ben generated by three sym­metry elements that are canonically chosen as $\mathcal{C}_6$, mirror $\mathcal{M}_{yz}$, inversion $\mathcal{P}$ and time-reversal $\mathcal{T}$. The Hamiltonian up to $O(\bm{k}^3)$ reads
\begin{eqnarray*}
  H(\bf{k})&=&\epsilon_0(\bf{k})+\left(\begin{array}{cccc}
      M(\bm{k}) & A(\bm{k})k_{+}& 0 & -B^*(\bf{k}) \\
 A(\bm{k})k_{-} &
   -M(\bm{k}) & B^*(\bf{k}) & 0
   \\
 0 & B(\bf{k}) & M(\bm{k}) &
   A(\bm{k})k_{-} \\
 -B(\bf{k}) & 0 & A(\bm{k})k_{+} & -M(\bm{k}) \\
\end{array}\right)
\end{eqnarray*}
where $\epsilon_0({\bf k})=C_{0}+C_{1}k_{z}^{2}+C_{2}(k_x^{2}+k_y^2)$,
$k_{\pm}=k_{x}\pm ik_{y}$, $A({\bf
  k})=A_{0}+A_{1}k_{z}^{2}+A_{2}(k_x^{2}+k_y^2)$, $B({\bf k})=B_3k_zk_{+}^2$ and $M({\bf
  k})=-M_{0}+M_{1}k_{z}^{2}+M_{2}(k_x^{2}+k_y^2)$,   with parameters $M_0,
M_1, M_2>0$ to reproduce band inversion.  The parameters for LiAuSe can be fitted by first-principles band dispersions as $C_0=0.0002$, $C_1=-0.0133$, $C_2=15.0321$, $A_0=0.4274$, $A_1=4.8621$, $A_2=145.701$, $M_0=0.1026$, $M_1=0.6818$, and $M_2=-0.0262$. Please note that under $C_6$ symmetry, the leading-term of off-diagonal should take the third-order term of $\bm{k}$. The $4\times4$ Hamiltonian could be diagonalized and the eigenvalues are $E({\bf k})=\epsilon_0({\bf k})\pm\sqrt{M({\bf
    k})^{2}+A^{2}k_+k_-+|B_3|^2k_z^2k_{+}^2k_{-}^2}$.  Then, we get the gapless points  at $\bm{k}_c=(0,0,\pm\sqrt{M_0/M_1})$,  which are the Dirac nodes along $\Gamma$-$A$.  In the vicinity of the gapless points $\bm{k}_c$, the leading-term of the solution is linear, and confirm that the gapless points are nothing but 3D massless Dirac points. 
    
We further write the effective $k\cdot p$ Hamiltonian   in a more concrete form as
\begin{equation}  
\begin{split} 
H_{DSM}(\bm{k})=  \epsilon_0(\bm{k})\sigma_0\tau_0 + M(\bm{k})\sigma_0\tau_z + A(\bm{k})(k_x\sigma_0\tau_x \\
- k_y\sigma_z\tau_y) +B_3k_z(k_x^2-k_y^2)\sigma_y\tau_y - 2B_3k_zk_xk_y\sigma_x\sigma_y
\end{split}
\end{equation}
where $\sigma_{x,y,z}$ and $\tau_{x,y,z}$ are Pauli matrices, and $\sigma_0$ and $\tau_0$ are unit matices for spin and orbital respectively.

For Weyl semimetal, the effective $k\cdot p$ Hamiltonian reads
\begin{equation}   
H_{WSM}(\bm{k})=H_{DSM}(\bm{k})+h\sigma_z\tau_z
\end{equation}
where $h$ describes the exchange field strength. After applying this term, the Dirac points will split into two separated Weyl points in momentum space, as shown in Fig.~\ref{fig4}. 

When the $C_6$ rotational symmetry is broken, the DSM transform into TI phase. A linear order of off-diagonal term will be introduced. Then the effective $k\cdot p$ Hamiltonian reads
\begin{equation}  
\begin{split} 
H_{TI}(\bm{k})=  \epsilon_0(\bm{k})\sigma_0\tau_0 + M(\bm{k})\sigma_0\tau_z + A(\bm{k})(k_x\sigma_0\tau_x - k_y\sigma_z \\
\tau_y)+B_3k_z(k_x^2-k_y^2)\sigma_y\tau_y - 2B_3k_zk_xk_y\sigma_x\sigma_y +B_1k_z\sigma_y\tau_y
\end{split}
\end{equation}
where the leading-term of off-diagonal becomes linear. Please note that this term $B_1k_z\sigma_y\tau_y$ is not unique. Any term that preserves all the above symmetries except $C_6$ is possible.  For convenience, we choose the $k_z$ term here. After introducing the linear term in off-diagonal elements, the system transform from DSM to TI phase.  The conclusion has been confirmed by our first-principles calculations (see Fig.~\ref{fig5}).

\section{Conclusion}

In summary, we have studied the topological phase transition from Dirac semimetal to Weyl semimetal or topological insulator by taking the ternary compound LiAuSe with hexagonal crystal structure as an example.   Both first-principles method and effective low energy $k\cdot p$ model were used in the calculations and result analysis. Our results show that Dirac semimetal is a critical phase point for topological phase transition. Since a $4\times 4$ Dirac Hamiltonian could be decomposed into  a pair of $2\times 2$ Weyl  nodes by breaking time reversal or inversion symmetries, it is not stable in general and additional crystal symmetry protection is necessary. Rotational symmetry plays a very important role in keeping the stability of the Dirac nodes at the symmetry axis for systems with time reversal symmetry. Thus, by breaking this term, the unstable system would transform from Dirac semimetal to topological insulator. If the rotational symmetry is kept but time-reversal symmetry is broken, it will transform from Dirac semimetal to Weyl semimetal. In LiAuSe material, we find WSM phase could be realized by Sm doping or applying external Zeeman field. Our work provides an excellent playground for the study of topological phase transition, and it is suitable to most topological semimetal with accidental Dirac points. SOC is considered in all the above results. It is interesting that we find a novel starfruit-like nodal chain semimetal state in LiAuSe in the absence of SOC. We expect that our predictions here could give a new  perspective on topological states and can be verified experimentally in the future.

\begin{acknowledgement}

The authors acknowledge helpful discussions with S. A. Yang, Z.-M Yu and H.-M. Guo. This work is supported by the NSF of China (No. 11504013, No.11474015, No.61227902 and No.11774018).

\end{acknowledgement}



\bibliography{draft_ref}





\end{document}